\documentclass[twocolumn,preprintnumbers,amsmath,amssymb]{revtex4}
\usepackage{graphicx}
\usepackage{dcolumn}
\usepackage{bm}
\usepackage[final]{pdfpages}
\begin{document}
\title{\textbf{Interplay between antiferrodistortive, ferroelectric and superconducting instabilities in Sr$_{1-x}$Ca$_{x}$TiO$_{3-\delta}$}}
\author{B. S. de Lima$^{1}$, M. S. da Luz$^{1,2}$, F. S. Oliveira$^{1}$, L. M. S. Alves$^{1}$, C. A. M. dos Santos$^{1}$,
F. Jomard$^{3}$,  Y. Sidis$^{4}$, P. Bourges$^{4}$, S.~Harms$^{5}$, C.P.~Grams$^{5}$, J. Hemberger$^{5}$, X. Lin$^{6}$, B. Fauqu\'e$^{6}$, and K. Behnia$^{6}$}
\affiliation{(1) Universidade de S$\tilde{a}$o Paulo, Escola de Engenharia de Lorena, Lorena, Brazil\\
(2) Universidade Federal do Tri\^angulo Mineiro, Instituto de Ci\^{e}ncias Tecnol\'{o}gicas e Exatas, Uberaba, Brazil\\
(3) Groupe d'\'etude de la mati\'ere condens\'ee, CNRS/UVSQ, 78035 Versailles, France\\
(4) Laboratoire L\'eon Brillouin, CEA-CNRS,  91191 Gif sur Yvette, France\\
(5) II. Physikalisches Institut, Universität zu K\"{o}ln,  50937 K\"{o}ln, Germany\\
(6) Laboratoire de Physique et Etude de Mat\'{e}riaux, CNRS/ESPCI/UPMC, 75005 Paris, France\\
}
\date{October 24, 2014}


\begin{abstract}
SrTiO$_{3}$ undergoes a cubic-to-tetragonal phase transition at 105K. This antiferrodistortive transition is believed to be in competition with incipient ferroelectricity. Substituting strontium by isovalent calcium induces a ferroelectric order. Introducing  mobile electrons to the system by chemical non-isovalent doping, on the other hand, leads to the emergence of a dilute metal with a superconducting ground state. The link between superconductivity and the other two instabilities is a  question gathering momentum in the context of a popular paradigm linking unconventional superconductors and quantum critical points. We present a set of specific-heat, neutron-scattering,dielectric permittivity and polarization measurements on Sr$_{1-x}$Ca$_{x}$TiO$_{3}$ ($0<x<0.009$) and a study of low-temperature electric conductivity in Sr$_{0.9978}$Ca$_{0.0022}$TiO$_{3-\delta}$. Calcium substitution was found to enhance the transition temperature for both anti-ferrodistortive and ferroelectric transitions. Moreover, we find that Sr$_{0.9978}$Ca$_{0.0022}$TiO$_{3-\delta}$ has a superconducting ground state. The critical temperature in this rare case of a superconductor with a ferroelectric parent, is slightly lower than in SrTiO$_{3-\delta}$ of comparable carrier concentration. A three-dimensional phase diagram for Sr$_{1-x}$Ca$_{x}$TiO$_{3-\delta}$ tracking the three transition temperatures as a function of x and $\delta$ results from this study, in which ferroelectric and superconducting ground states are not immediate neighbours.
\end{abstract}

\maketitle

\begin{center}
\textbf{I. Introduction}
\end{center}

SrTiO$_{3}$ is a large-gap semiconductor, belonging with the ABO$_{3}$ family of perovskites. At room temperature, it is  cubic, filling the space  with an interposition of TiO$_{6}$ octahedra and SrO$_{12}$ cuboctahedra. Upon cooling, it undergoes a structural cubic-to-tegragonal phase transition below 105 K. This transition and in particular its soft mode and its non-classical exponents were subject to numerous studies\cite{Cowley}. The tetragonal distortion, leading to the $c/a$ ratio becoming slightly  ($5.6 \times 10^{-4}$) larger than unity\cite{Lytle}, is accompanied by rotation of the TiO$_{6}$ octahedra around one of the three tetragonal axes. The angle of rotation saturates to 2.1 degrees at low temperature\cite{Unoki,Muller1}. This antiferrodistortive transition (an intriguing case of ferroelasticity\cite{Salje}), continues to attract significant attention\cite{Salje1}.

\begin{figure} [htp]
\includegraphics[width=8cm]{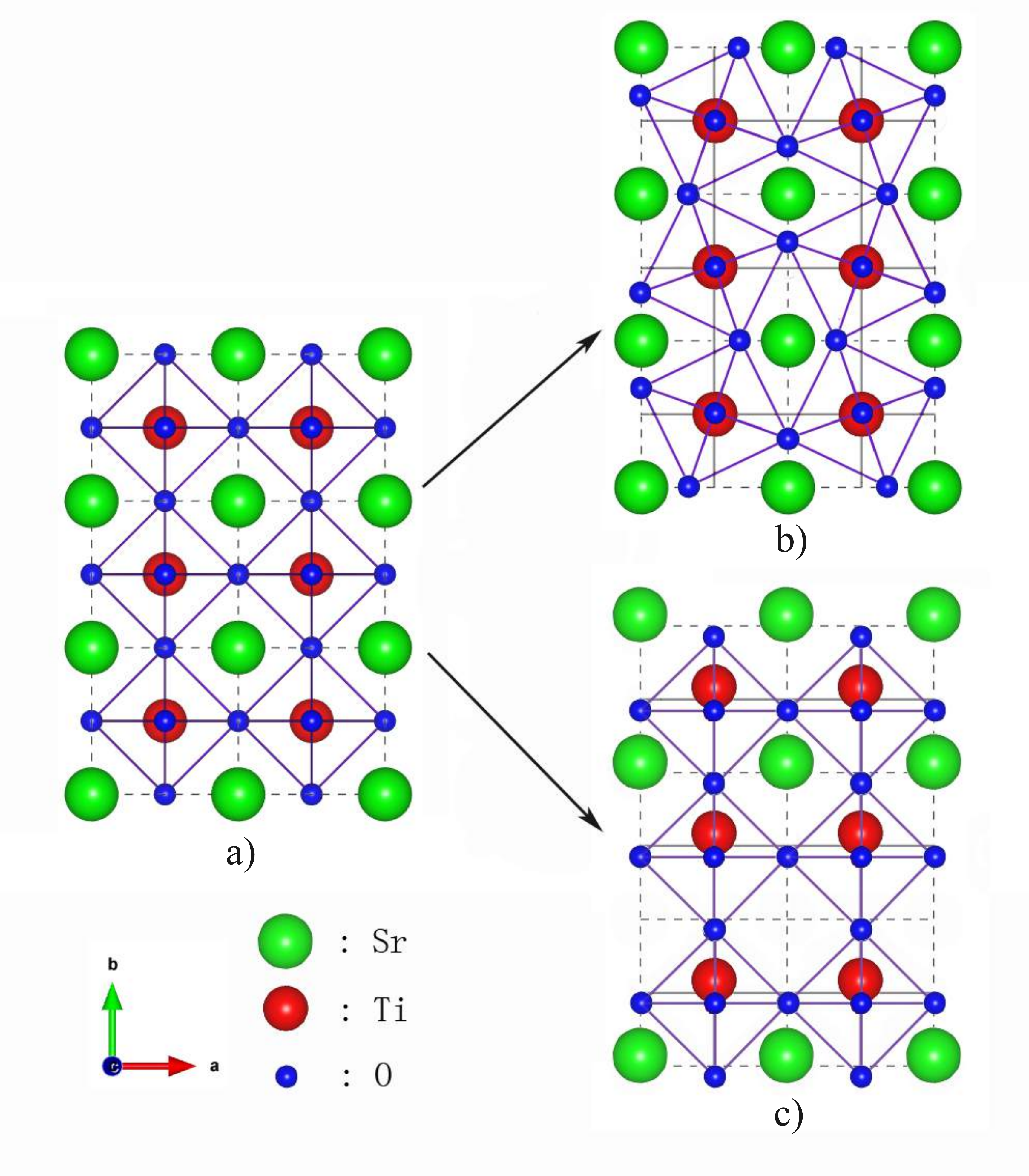}
\includegraphics[width=8cm]{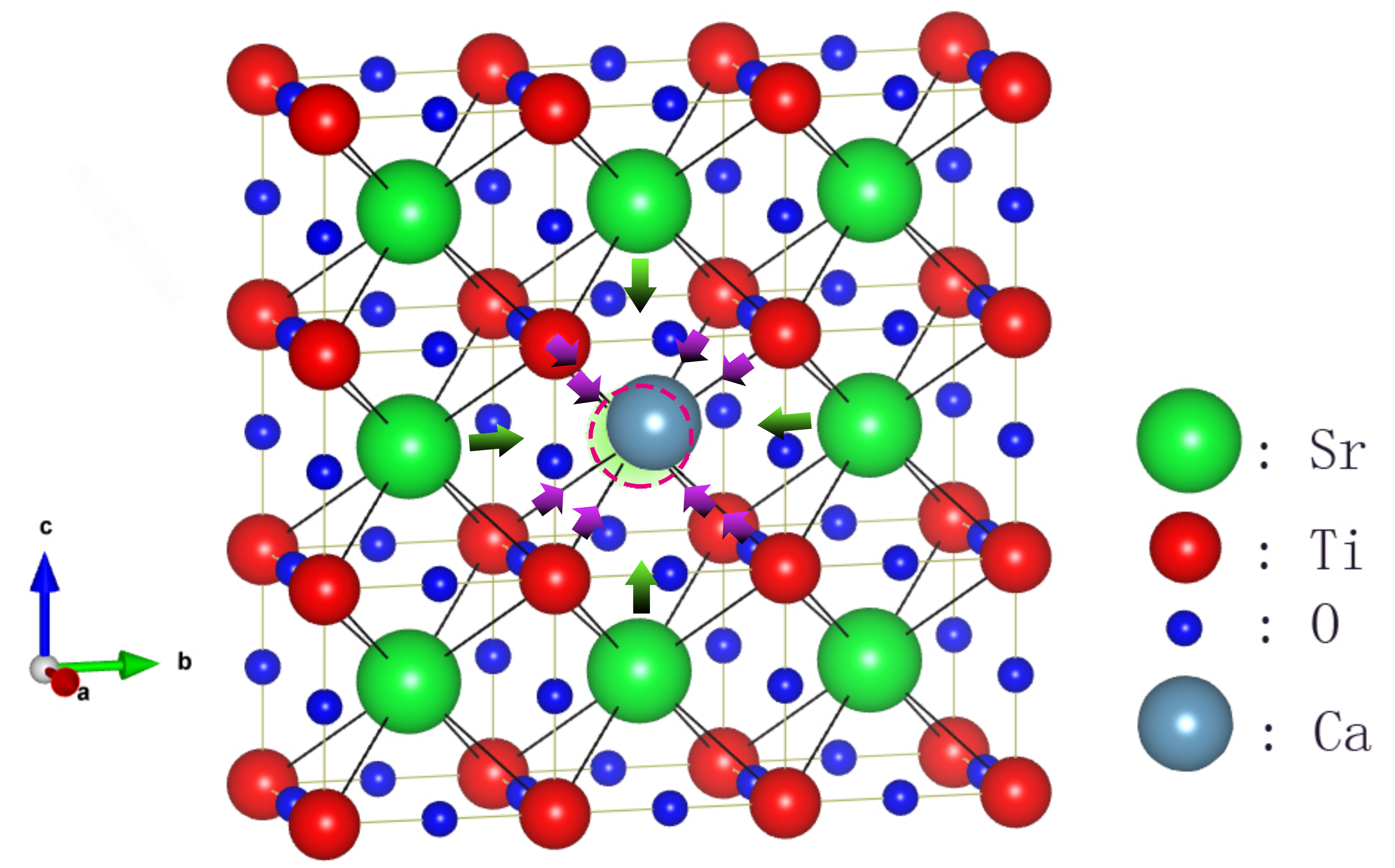}
\caption{Top: Crystal structure of SrTiO$_{3}$ seen along the (100) axis (a). The lattice can be may distorted by two instabilities, antiferrodistortive (b) and ferroelectric(c) . Bottom: Crystal structure of Ca-doped SrTiO$_{3}$. By replacing larger strontium, calcium finds itself at a slightly off-the-center position of a CaO$_{12 }$ cubocathedron. The substitution creates a local electric dipole and a strain field shown by arrows. Percolation of dipoles leads to ferroelectricity, while the strain field strengthens the antiferrodistortive transition.}
\end{figure}

SrTiO$_{3}$ is also a quantum paraelectric\cite{Muller2}. In contrast to several members of the ABO$_{3}$ family, such as BaTiO$_{3}$ or PbTiO$_{3}$, which become ferroelectric, it avoids this instability thanks to large quantum fluctuations\cite{Muller2,Hemberger}. The proximity to ferroelectricity leads to a large dielectric constant, which saturates to an amplitude exceeding the typical values for insulating transition metal oxides by several orders of magnitude\cite{Muller2}. The competitive or cooperative interplay between the ferroelectric and antiferrodistortive distortions has been a subject of theoretical attention\cite{Zhong,Benedek,Aschauer}.

One can introduce electron-like carriers to this insulator by doping it. This can be done either by substituting titanium by niobium, or strontium by lanthanum or by simply removing oxygen. In 1964, it was discovered that this n-doped SrTiO$_{3}$ has a superconducting ground state\cite{Schooley}. Intriguingly, superconductivity is limited to a narrow range of carrier concentration\cite{Schooley2,Koonce,Binnig, Lin1,Lin2}. It starts at a carrier density as low as 10$^{-5}$ electron per formula unit\cite{Lin1} and ends when it exceeds 0.02 electron per formula unit\cite{Koonce,Suzuki}. The low level of carrier concentration at which this superconducting dome emerges is in singular contrast to all other known cases of superconductors with an insulating parent. It is tempting to relate it to the precocious metal-insulator transition triggered by the long effective Bohr radius, which becomes as long as a fraction of micron\cite{Lin1,Spinelli} in this system. The drastic enhancement of the Bohr radius is, in turn, a consequence of the large dielectric coefficient due to the proximity of a ferroelectric instability.

Decades ago, Bednorz and M\"uller discovered that substituting strontium with isovalent calcium leads to a ferroelectric ground state in  Sr$_{1-x}$Ca$_{x}$TiO$_{3}$ as soon as x exceeds 0.002\cite{Bednorz}. This was followed by several studies on the effect of doping by calcium on various physical properties\cite{Kleemann1,Bianchi,Kleemann2}. One interesting outcome of these studies was the observation of an enhancement in the low-temperature dielectric constant with Ca doping\cite{Bednorz,Kleemann2}. This implies that, at least for small $x$, the Bohr radius in Sr$_{1-x}$Ca$_{x}$TiO$_{3}$ is larger than in SrTiO$_{3}$. In this context, one may wonder how this would affect the threshold of both the metal-insulator transition and the emergent superconductivity in n-doped Sr$_{1-x}$Ca$_{x}$TiO$_{3}$.

In this paper we report on a study of  Sr$_{1-x}$Ca$_{x}$TiO$_{3}$  ($0<x<0.0091$) and Sr$_{0.9978}$Ca$_{0.0022}$TiO$_{3-\delta}$. We find that, according to the evolution of the associated specific heat anomaly, the antiferrodistortive transition temperature rapidly rises with calcium substitution. Neutron-scattering measurements confirm this result and indicate that calcium substitution  amplifies the rotation of TiO$_{6}$ octahedra. Measurements of dielectric constant confirm the emergence of ferroelectricity, most probably through the percolation of the polar Ca-substituted unit cells. This points to a cooperation between the two distortions in this doping window. Finally,  dilute superconductivity is observed in Sr$_{0.9978}$Ca$_{0.0022}$TiO$_{3-\delta}$ single crystals. In contrast to previous studies on ceramics\cite{Schooley3,Eagles}, these samples are single crystals and show quantum oscillations in presence of moderate magnetic fields leading to an accurate quantification of the concentration of mobile electrons.

In Sr$_{0.9978}$Ca$_{0.0022}$TiO$_{3-\delta}$, the superconducting critical temperatures were found to be slightly lower and the superconducting transitions somewhat broader than in SrTiO$_{3-\delta}$ crystals of comparable carrier concentrations. Two possibilities are discussed. Either oxygen vacancies cluster around calcium sites generating additional inhomogeneity or the lack of inversion symmetry at calcium-substituted sites weakens superconductivity.

Several theoretical proposals have linked dilute superconductivity to elementary excitations of the two neighboring orders. Our study is a first attempt to map an unusual case of vicinity between these three instabilities. A three-dimensional phase diagram for Sr$_{1-x}$Ca$_{x}$TiO$_{3-\delta}$ can be drawn.

\begin{figure} [htp]
\includegraphics[width=8cm]{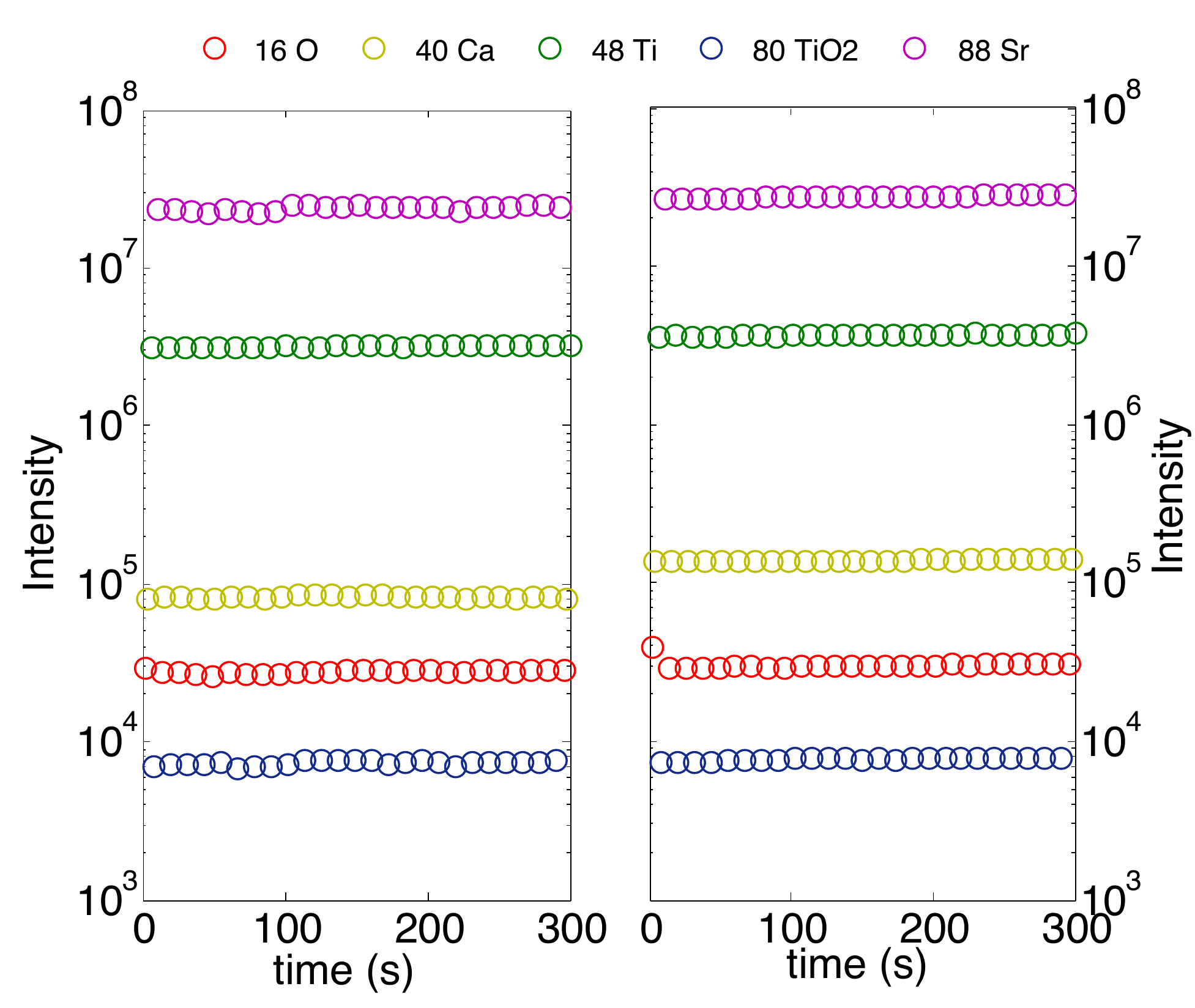}
\caption{ Secondary Ion Mass spectroscopic counts of ionic masses 16[oxygen], 40[calcium], 48[titanium] , 80[TiO2] and  88[strontium] in two Sr$_{1-x}$Ca$_{x}$TiO$_{3}$ single crystals used in this study, with x=0.0022 (left) and x=0.0045 (right). As seen in the figure,  the intensity of all other masses are identical in the two samples, except for calcium.}
\end{figure}

\begin{center}
\textbf{II. Experimental details}
\end{center}

Commercial SrTiO$_{3}$:Ca single crystals (provided by Crystec GMBH) were used in this study. Specific heat measurements were performed using a Quantum Design PPMS heat capacity module in a temperature window close to the reported temperature of the structural transition.

The dielectric measurements were performed from room temperature to 2~K with a home-made coaxial-line inset inserted in a PPMS cryostat. Therefore the samples were prepared as plate-capacitors with typical dimensions of $0.5\times 2.5\times 5$\,mm$^3$ and electrodes made of silver paint. Real and imaginary components of the frequency-dependent dielectric response were measured using a frequency-response analyzer (Novocontrol) in the frequency range between 1~Hz and 10~kHz with a stimulus of the order $E_{ac} \approx 1$~V/mm. In addition, electric field dependent polarization data in higher fields up to 300~V/mm (Novocontrol HVB1000) were obtained evaluating the non-linear permittivity contributions as described in [\onlinecite{Niermann14}]. Since no special surface etching was done after cutting the samples in order to remove eventual passivated surface-layers\cite{Bednorz,Kleemann1}, the absolute values of the of the dielectric measurements should be taken with caution.

Neutron diffraction data were collected on the 3T1 diffractometer at the Orph\'ee reactor in Laboratoire L\'eon Brillouin (Saclay, France) with an incident neutron beam at E$_i$  =14.7 meV and a 20' collimation on the scattered beam.  Two PG filters were inserted on the scattered beam in order to eliminate double scattering. The crystal had a volume of 25mm$^3$ and a resolution-limited mosaicity of 0.2$^\circ$.

Oxygen-deficient  samples  were obtained by annealing in a temperature range of $700-1000$ $^\circ$C in a vacuum of $1 10^{-6}$ mbar for 1 to 2 hours. Ohmic contacts were obtained by evaporating gold on the samples and heat up to 550 $^\circ$C to promote gold diffusion into the crystals. Low-temperature longitudinal and Hall resistivity were  measured in a dilution system inserted in a 17 T superconducting magnet.

The nominal calcium concentration was cross-checked using the Secondary Ion Mass Spectrometry (SIMS) analysis technique, using IMS Cameca 7f equipment with primary oxygen beam. This technique allows to detect very low concentrations of dopants and to measure their in-depth distribution over a few microns.

Fig.~2 shows a typical profile at five atomic masses for two Sr$_{1-x}$Ca$_{x}$TiO$_{3}$ samples with $x=0.22\%$ and $x=0.45\%$ . The intensities are constant as a function of time, indicating that the concentration of each element is constant in the direction perpendicular to the surface.

The intensity at the four reference masses are comparable, but differ by a factor somewhat lower than two for the mass of calcium. By comparing the relative intensities in the two samples, one can obtain an accurate estimate of the relative concentration of a given element in them. The relative Ca/Sr ratio in the two samples differ by a a factor of 1.5$\pm$0.1 compared to a nominally expected value of 2. This sets the limit of our accuracy in the calcium content of these samples.

\begin{center}
\textbf{III. Antiferrodistortive order}
\end{center}

\begin{figure} [htp]
\includegraphics[width=8cm]{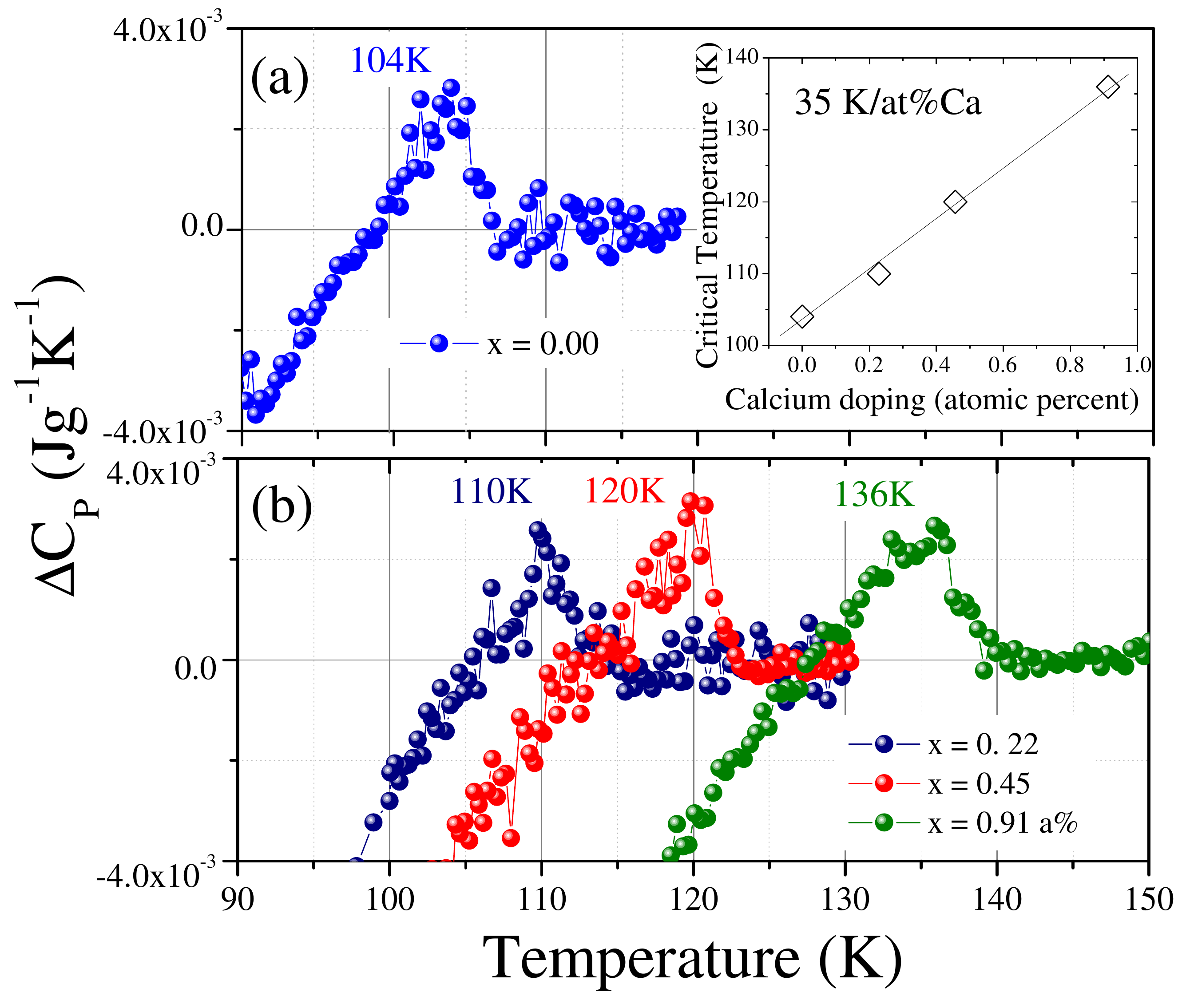}
\caption{The specific heat jump caused by the antiferrodistortive phase transition in SrTiO$_{3}$(top) and Sr$_{1-x}$Ca$_{x}$TiO$_{3}$(bottom). The inset of upper panel shows the evolution of the transition temperature with doping.}
\end{figure}

The structural phase transition in strontium titanate has been subject to numerous studies. Salje and collaborators\cite{Salje2} were the first to carefully resolve the excess specific heat associated with this phase transition. Based on their data, they argued that the transition is nearly tricritical and mean-field. We performed specific heat measurements on our samples and detected the excess specific heat caused by the antoferrodistortive transition.
The data for pristine SrTiO$_{3}$ is shown in the upper panel of Fig.3 . The amplitude of the anomaly is small. The jump in C/T is 3 mJg$^{-1}$K$^{-1}$, barely more than one percent of the overall background signal. We extracted it by subtracting a polynomial fit to the background signal in a manner similar to Salje \emph{et al.}\cite{Salje2}. Both the magnitude and the shape of the anomaly are  similar to what was found in several previous reports\cite{Salje2,Salje3,Bussmann}.

We did similar measurements on three Ca-doped samples with different doping levels. The results can be found in the lower panel of the figure. Sharp specific heat anomalies indicate homogeneous Ca distribution in the samples. One can clearly see that the specific heat anomaly shifts to higher temperatures as Ca doping increases. As can be seen in the inset of the upper panel, the shift corresponds to linear increase of 35K/\%Ca. To the best of our knowledge,
this is the first set of specific heat measurements showing the evolution of the structural phase transition in the Sr$_{1-x}$Ca$_{x}$TiO$_{3}$ system. We also measured the specific heat over a large temperature interval below the structural phase transition. Within experimental resolution, we did not find any other anomaly. A similar conclusion was reported by a previous specific-heat study on pristine SrTiO$_{3}$\cite{Salje3}.

We also performed neutron diffraction measurements to probe this phase transition. The crystal structure becomes I4/mcm tetragonal below 105 K. In a pseudo cubic indexing, the structural transition gives rise to new Bragg reflections (h,k,l) with h,k,l half integers such as {\bf{Q}}=(1.5,0.5,0.5).

\begin{figure}[htbp]
{\includegraphics[width=8cm]{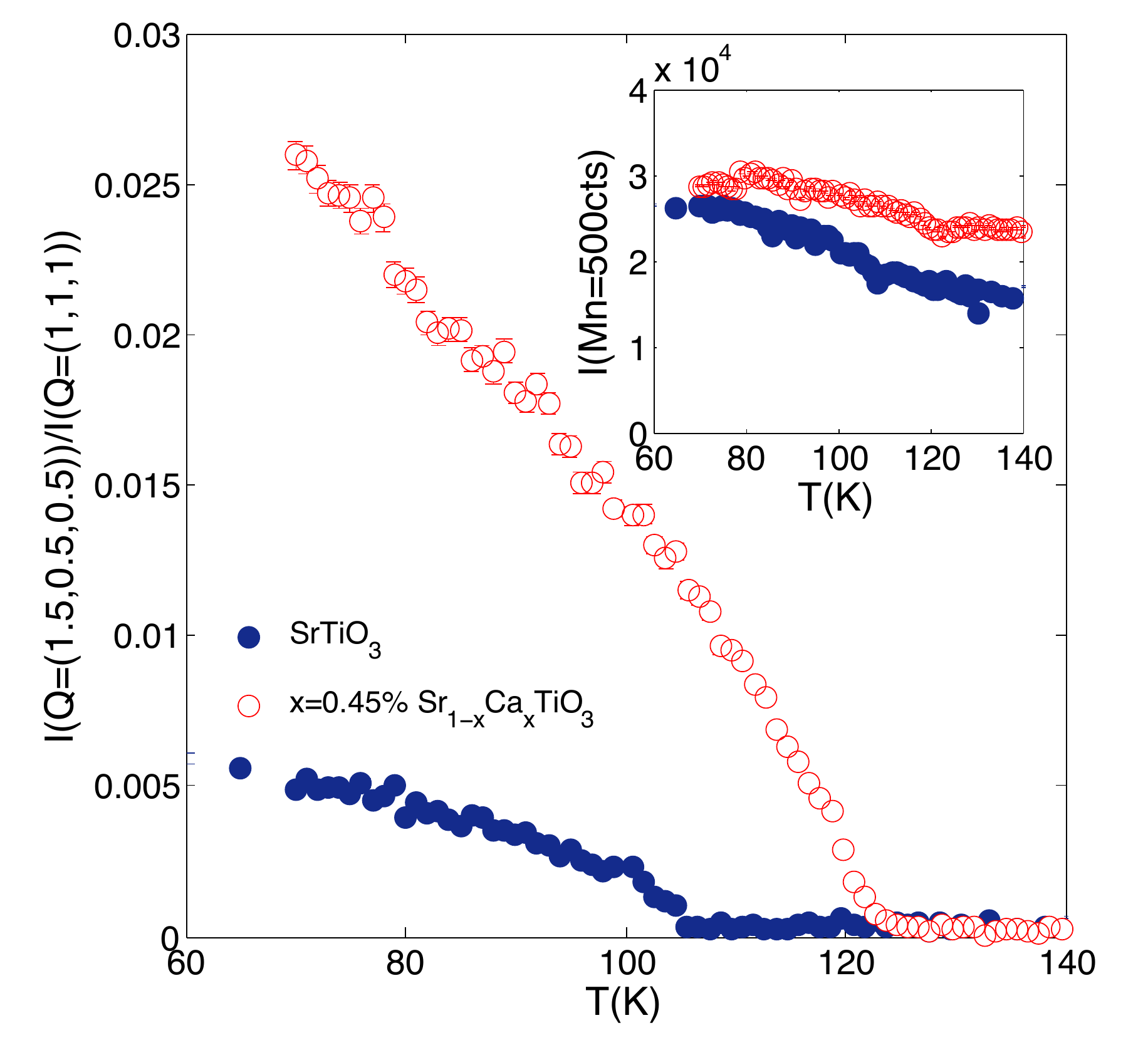}}\caption{Temperature dependence of the normalized intensity at {\bf{Q}}=(1.5,0.5,0.5) for SrTiO$_3$ (blue point) and  Sr$_{1-x}$Ca$_x$TiO$_3$ x=0.0045. The inset shows the temperature dependence of the intensity of the  {\bf{Q}}=(1,1,1) Bragg peak.}
\label{Neutron}
\end{figure}

Fig. 4 compares the temperature dependence of the intensity at {\bf{Q}}=(1.5,0.5,0.5) normalized by the intensity at {\bf{Q}}=(1,1,1) of Sr$_{1-x}$Ca$_x$TiO$_3$ in a pristine ($x=0$) and a doped ($x=0.00045$) sample. In both cases, the normalized intensity increases below a temperature, which was found to be T=106~K for $x=0$ and T=122~K for $x=0.0045$, in very good agreement with the specific heat results. In addition, we note that the normalized intensity is five times higher for the Ca-doped sample than for the undoped sample. As seen in the inset, the intensity of the Bragg peak {\bf{Q}}=(1,1,1) for both samples is little affected by the structural transition and displays the same temperature dependence. Therefore, the difference in the normalized intensities can be safely attributed to a change in the structural factor in the Ca-doped compound.

According to the early work by Shirane and co-workers\cite{Shirane}, the structural factor (F) of the superlattice peak depends only on the rotation of the octahedara. It can be written as :$F=16\pi(h+k)\delta$, where $\delta$=$1/4 \sin (\varphi)$ and $\varphi$ is the angle of rotation of the octahedra. In the case of pure SrTiO$_3$, electron spin resonance \cite{Unoki} and neutron scattering measurements \cite{Shirane} find $\varphi =1.4^\circ$ at 70~K. Therefore, assuming an unchanged symmetry, the higher intensity of the {\bf{Q}}=(1.5,0.5,0.5) Bragg peak  in Ca-doped SrTiO$_3$ would  suggest that the octahedra tilt is enhanced by a factor of 2.3, meaning that $\varphi =3.2^\circ$ at 70~K. In other words, calcium doping not only shifts the transition to higher temperature, it also enhances the rotation angle.

Thus, two distinct experimental probes, specific heat and neutron scattering, find that calcium doping strengthens the structural transition. We notice that this result is in agreement with the phase diagram drawn by Ranjan\cite{Ranjan} scrutinizing a number of previous reports by other authors.

\begin{center}
\textbf{III. Ferroelectricity}
\end{center}

The temperature-dependence of real and imaginary components of the permittivity $\varepsilon^*(T)$ of the Sr$_{1-x}$Ca$_{x}$TiO$_{3}$ samples were measured from 300~K to 2~K and for Ca concentrations of $x=0,~ 0.22\%,~0.45\%,~\mathrm{and}~0.91\%$. The results are presented in Fig.~5.

\begin{figure} [htp]
\includegraphics[width=0.49\textwidth]{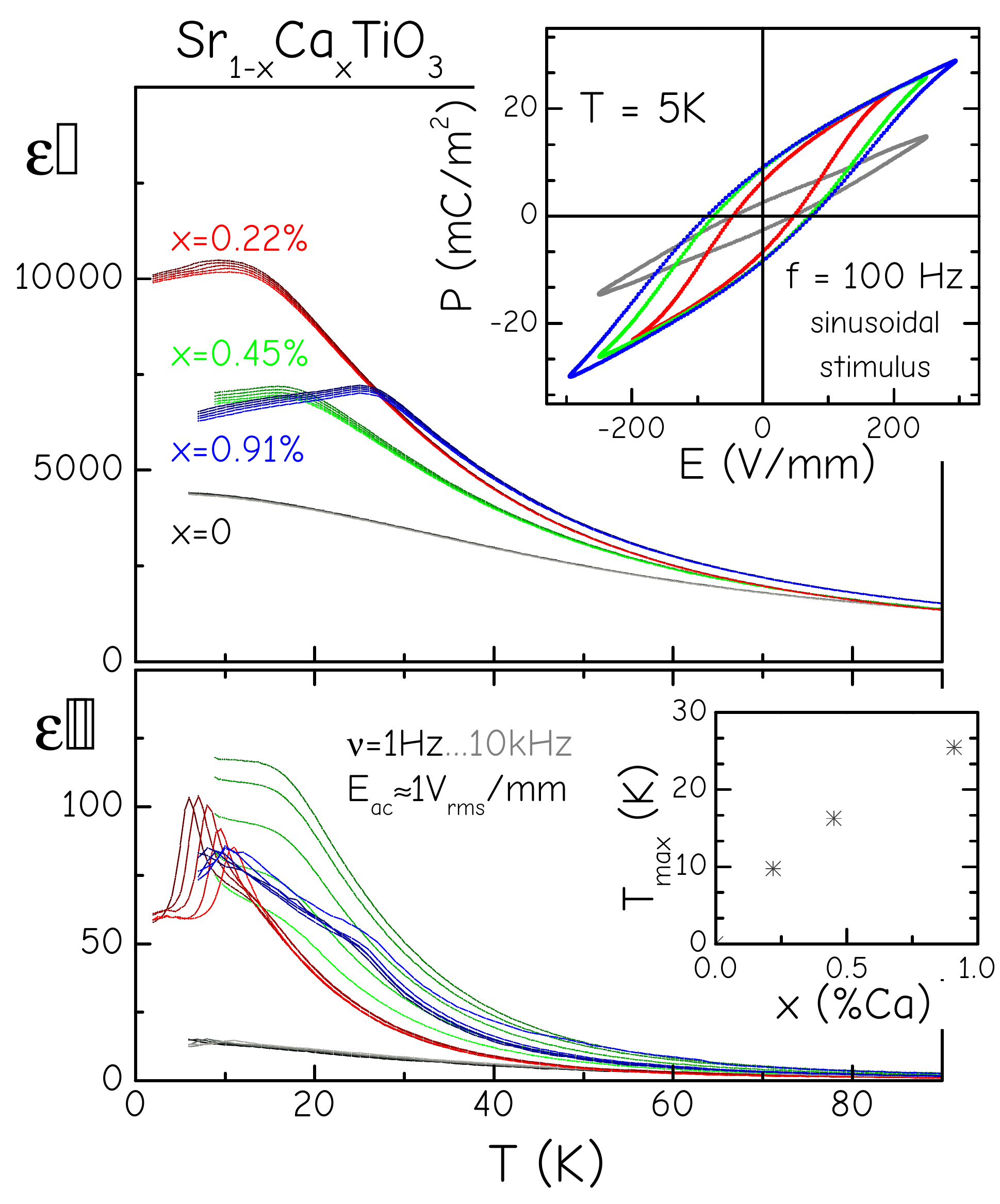}
\caption{Temperature dependence of the real and imaginary components of the dielectric permittivity in Sr$_{1-x}$Ca$_{x}$TiO$_{3}$. The inset show $P(E)$ hysteresis loops at $T=5$~K.}
\end{figure}

Pure STO ($x=0$) displays the well-known quantum-paraelectric behaviour: The real part $\varepsilon'(T)$ increases with decreasing temperature and saturates at low temperatures ($T\leq 40$~K) where the potential ferroelectric order is suppressed due to quantum fluctuations. The dielectric loss $\varepsilon''(T)$ shows a similar curvature but roughly three orders of magnitude smaller. Within the explored frequency range (1~Hz~$<$~$\nu$~$<$1~kHz) no pronounced dispersive features can be found in the pure sample.

Introducing a Ca-concentration of $0.22\%$ leads to a change in the curvature of the temperature-dependent permittivity.
The concentration $x=0.22\%$ is right on the edge between a quantum paraelectric ground state like in pure STO and ferroelectric order but already exhibits a smeared-out maximum in $\varepsilon'(T)$ around $T_{max}\approx 10$~K. The loss $\varepsilon'(T)$ shows additional contributions compared to the pure compound below $T\approx50K$ and below $T_{max}$ dispersive loss peaks can be found, shifting to lower temperatures with decreasing frequency. These loss features can be understood by the formation of polar nano-regions seeded by the electric dipoles of off-center calcium sites embedded in the highly polarizable STO lattice. Even if these clusters remain disordered at low $Ca$-concentrations they may experience glass-like freezing at low temperatures\cite{Kleemann3} or they may give rise to additional loss at higher temperatures due to interaction with eventually polar structural domain walls arising below the tetragonal phase transition \cite{Viana,Salje1}.

For the higher concentrations ($x=0.0045$ and $x=0.0091$), the maximum in $\varepsilon'(T)$ and the dispersive features in $\varepsilon''(T)$ both shift to higher temperatures. The ferroelectric transition in Sr$_{1-x}$Ca$_x$TiO$_3$ is believed to be percolative. At high-enough calcium concentration, the Ca-induced nano-clusters begin to interact and at low-enough temperature an order emerges. As displayed in the lower inset of Fig.\,5 the temperature $T_{max}$ for which $\varepsilon'(T)$ exhibits a maximum shift upwards with calcium doping.

The shape of these maxima in $\varepsilon'(T)$ is flatter compared to canonical ferroelectrics, which show divergent behavior and a vanishing contribution to the permittivity for lowest temperatures. The temperature dependent permittivities in this concentration range of Sr$_{1-x}$Ca$_x$TiO$_3$ seem to extrapolate to a finite value for $T\approx 0$, denoting the influence of quantum fluctuations. Therefore, these materials are referred to as quantum ferroelectrics \cite{Kleemann3,Bednorz}.

As the upper inset of Fig.~5 shows, the $P(E)$-loops at low temperatures in the Ca-doped samples reveal the typical non-linear characteristics of switchable remnant polarization as expected for ferroelectric materials. However, for high electric field these $P(E)$-loops do not flaten into a saturation value for the spontaneous polarization but keep a high slope corresponding to a finite value for the permittivity.
This aspect of polar (quantum) fluctuations and concomitantly high permittivity values persisting towards lowest temperatures and high fields may be of importance with respect to the onset of metallicity and even superconductivity due to charge carrier doping, as discussed in the following.

\begin{center}
\textbf{IV. Metallicity and superconductivity}
\end{center}
\begin{figure} [htp]
\includegraphics[width=8.5cm]{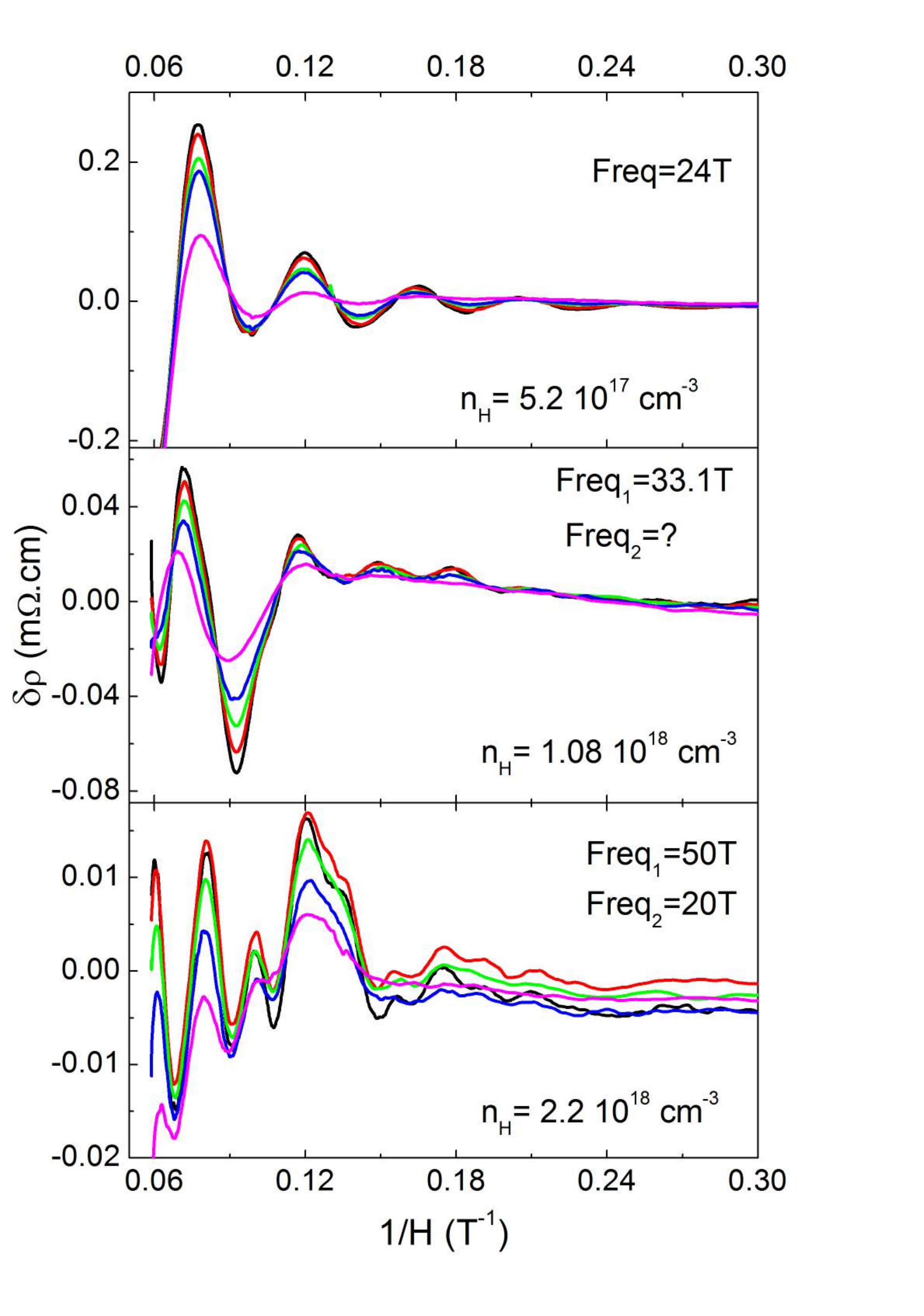}
\caption{\label{Fig.6}Shubnikov-de Haas oscillations seen in magnetoresistance of Sr$_{1-x}$Ca$_{x}$TiO$_{3}$ ($x= 0.22 at\%$) with carrier concentrations of \textit{n} = 5.2 10$^{17}$cm$^{-3}$ (top), \textit{n} = 1.08 10$^{18}$cm$^{-3}$ (middle) and \textit{n} = 2.2 10$^{18}$cm$^{-3}$ (bottom) panels. In each panel each curve corresponds to a different temperature.}
\end{figure}

Up to here, we have presented our data acquired by studying an insulating system with no mobile electrons. One can remove a tiny fraction of oxygen atoms out of Sr$_{0.9978}$Ca$_{0.0022}$TiO$_{3}$ by annealing samples samples in vacuum. As in the case of  SrTiO$_{3}$ with no calcium\cite{Spinelli}, we found that an extremely small concentration of oxygen vacancies is sufficient for them to host mobile electrons at very low temperature. With a carrier density of the order of 10$^{18}$cm$^{-3}$, the low-temperature electric resistivity becomes as low as 1 m$\Omega$ cm, indicating a mobility in the range of 6000~cm$^{2}$V$^{-1}$s$^{-1}$.

The system has not only a finite conductivity, but also it qualifies as a metal in a stronger sense of the term. It has a sharp Fermi surface and there is a well-defined Fermi-Dirac distribution for all carriers. This is attested by Shubnikov-de Haas oscillations seen in magnetoresistance. Thi is shown in Fig. 6. The sample with the lowest carrier density($n=5.2\times10^{17}$cm$^{-3}$ according to the Hall coefficient) showed a single frequency of 24T. Assuming a spherical Fermi surface, this corresponds to a carrier density of $6.6\times10^{17}$cm$^{-3}$. Given the moderate anisotropy of the Fermi surface\cite{Uwe,Allen,Lin2}, the agreement between the two estimations of the carrier density is reasonable.

As the carrier density increases, quantum oscillations begin to present a structure and they present more than one frequency. In the case of SrTiO$_{3-\delta}$, the first critical doping, n$_{c1}$, above which two distinct frequencies are detectable was found to be about $1.2 \times10^{18 }$cm$^{-3}$ \cite{Allen,Lin2}. In the case of Sr$_{0.9978}$Ca$_{0.0022}$TiO$_{3-\delta}$, we clearly see a single frequency when the carrier density is $5.2\times10^{17}$cm$^{-3}$ and two distinct and quantifiable frequencies when it is $2.2\times 10^{18}$cm$^{-3}$. The sample with the carrier density of $ 1.08 \times10^{18}$cm$^{-3}$ seems to be very close to the critical doping. We conclude that n$_{c1}$, the threshold doping for the occupation of a new band has not significantly shifted with slight calcium substitution.

Measurements of low-temperature zero-field electrical resistivity revealed a superconducting transition. This is illustrated in Fig. 7. As seen in the figure, compared to SrTiO$_{3-\delta}$ of comparable carrier concentration, the superconducting transition becomes broader in Ca-doped samples and the critical temperature is somewhat lower.

\begin{figure} [htp]
\includegraphics[width=9.5cm]{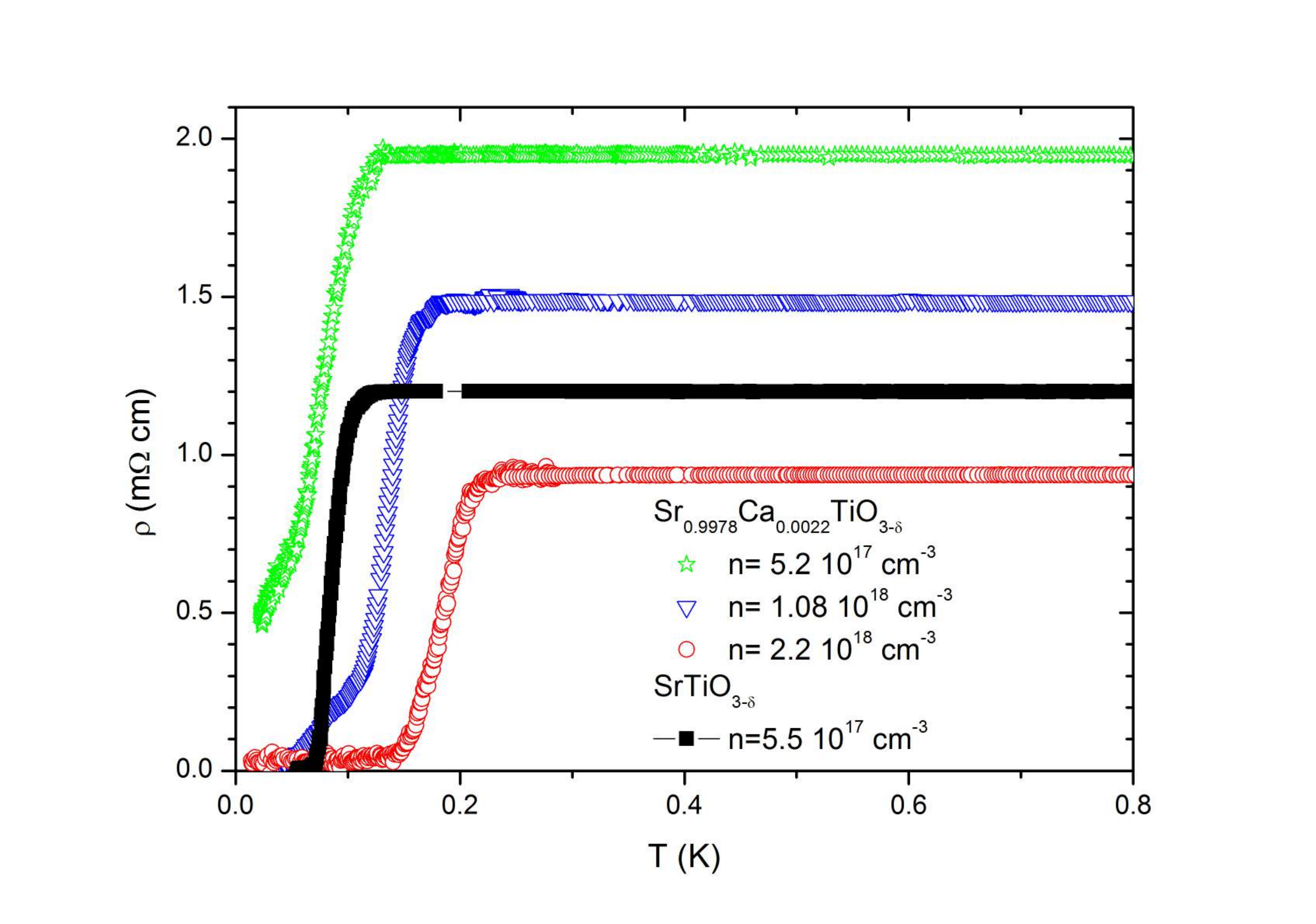}
\includegraphics[width=9.5cm]{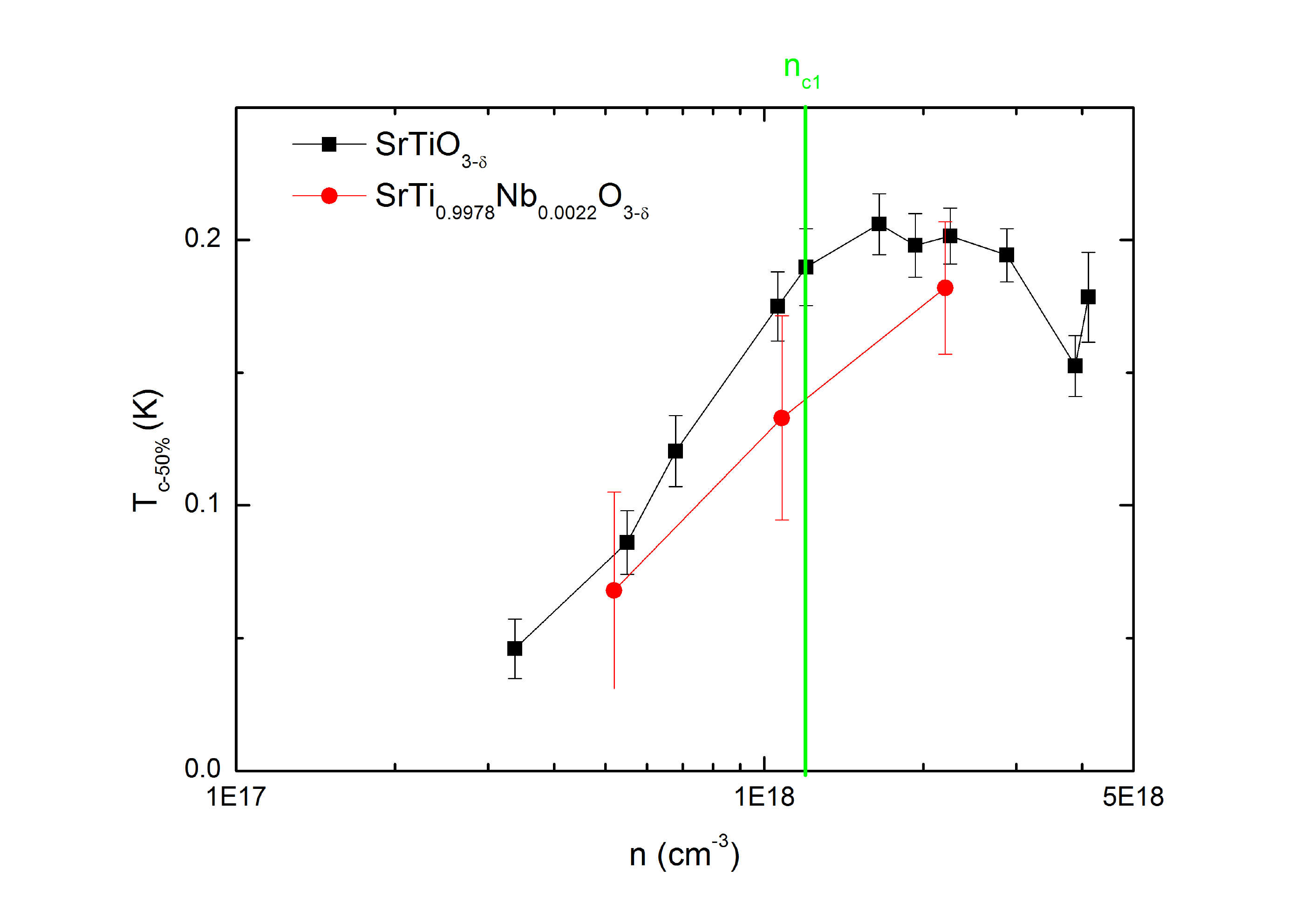}
\caption{Top: Superconducting resistive transitions in in Sr$_{0.9978}$Ca$_{0.0022}$TiO$_{3-\delta}$. A similar curve for  SrTiO$_{3-\delta}$  is shown for comparison. Bottom: Resistive critical temperature as a function of carrier concentration in Sr$_{0.9978}$Ca$_{0.0022}$TiO$_{3-\delta}$ and SrTiO$_{3-\delta}$. }
\end{figure}
\begin{figure} [htp]
\includegraphics[width=8.5cm]{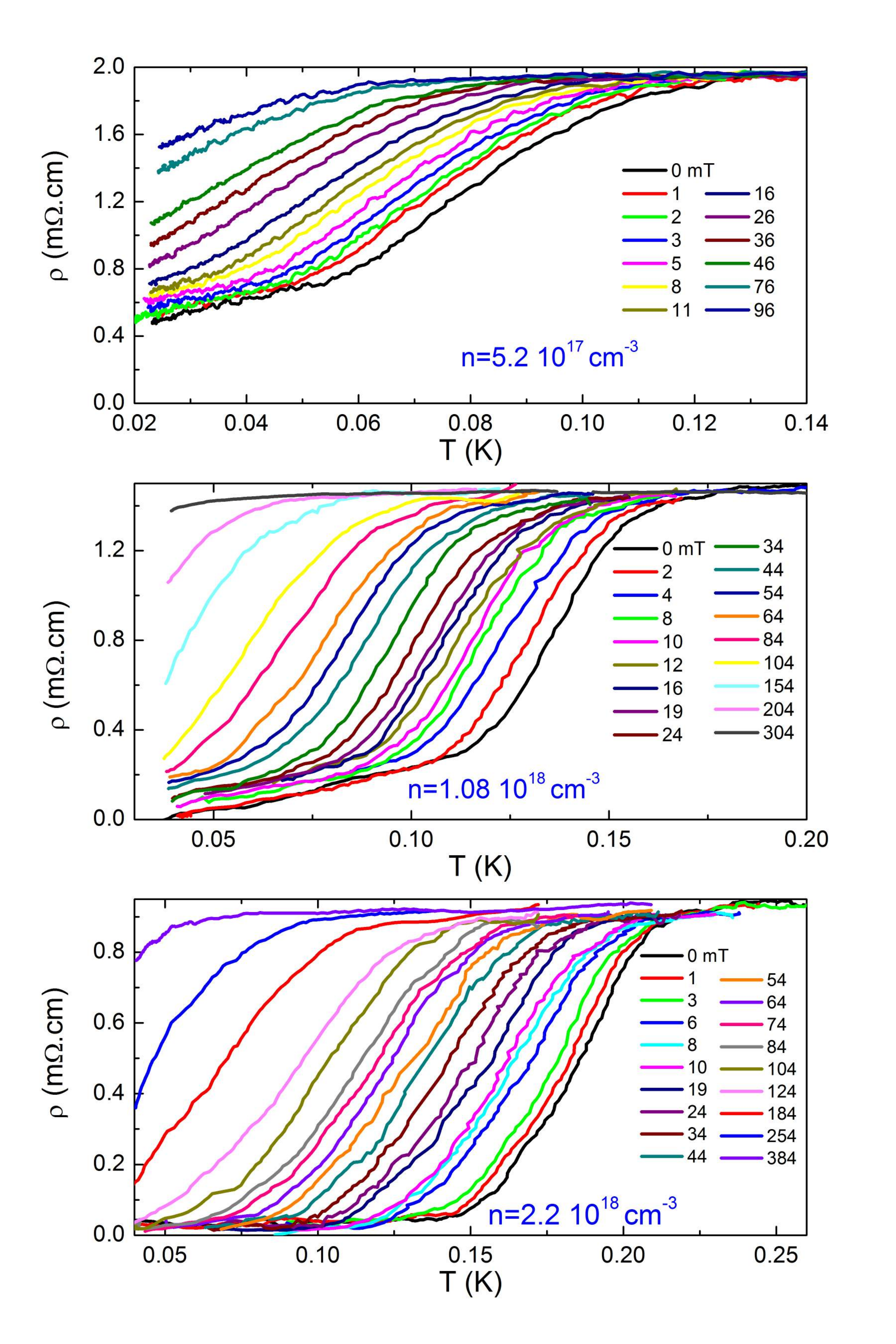}
\caption{Electrical resistivity as a function of temperature in Sr$_{0.9978}$Ca$_{0.0022}$TiO$_{3-\delta}$ for samples with different carrier densities; Top: $n = 5.2\times10^{17}$cm$^{-3}$; Middle: $n = 1.08 \times10^{18}$cm$^{-3}$ and bottom: $n = 2.2 \times10^{18}$cm$^{-3}$ at several applied magnetic fields.}
\end{figure}

The resistive transitions in these samples were studied in presence of magnetic field, as seen in Fig. 8. The transitions shift to lower temperature with increasing magnetic field. A resistive upper critical field can be extracted from the data. One can use the slope of the upper critical field near the critical temperature to quantify the superconducting coherence length. Note however, that according to a recent study of bulk Nb-doped SrTiO$_{3}$\cite{Lin3}, bulk probes of superconductivity yield a critical temperature and an upper critical field  significantly lower than what is obtained by resistivity measurements. Therefore, the values obtained here should be taken cautiously.

Table I gives a brief account of metallic and superconducting properties of our Sr$_{0.9978}$Ca$_{0.0022}$TiO$_{3-\delta}$ single crystals. The superconducting coherence length has been estimated, using $\xi^{-2}=\frac{2\pi}{\Phi_{0}}0.69T_{c}\frac{dH_{c2}}{dT}|_{T_{c}}$. The Dingle mobility was estimated by a Dingle fit to the SdH data. The Hall mobility was quantified by taking the ratio of Hall-to-longitudinal resistivity. As seen in the table, the normal state of is dirtier in Sr$_{0.9978}$Ca$_{0.0022}$TiO$_{3-\delta}$ than in SrTiO$_{3-\delta}$ and a comparison of the mean-free-path and the superconducting coherence length suggests that the system has become a dirty superconductor upon calcium doping.

\begin{table}[htp]
\caption{Properties of Sr$_{0.9978}$Ca$_{0.0022}$TiO$_{3-\delta}$ single crystals. Dingle $\mu_{D}$ and Hall $\mu_{H}$ mobilities as well as the mean-free-path $mfp$ , extracted from Hall mobility, and superconducting coherence length $\xi$ extracted from the slope of the upper critical field near T$_{c}$ are given. Values for SrTiO$_{3-\delta}$) samples of comparable concentration are given in parenthesis.} 
\centering 
\begin{tabular}{c c c c c} 
\hline\hline 
\textit{n} & $\mu_{H}$ & $\mu_{D}$ & $mfp$ & $\xi$\\ [0.5ex] 
cm$^{-3}$ & cm$^{2}V^{-1}s^{-1}$ & cm$^{2}V^{-1}s^{-1}$& nm & nm\\ [0.5ex] 
\hline 
5.2 10$^{17}$ & 6100 (9000) & 1530 (2500) & 108 (138) & 374 (99) \\ 
1.08 10$^{18}$ & 3900 (10000) & 1130 (3300) & 81 (186) & 108 (79) \\
2.2 10$^{18}$ & 3000 (8000) & --- & 80 (180) & 84 (58) \\ [1ex] 
\hline 
\hline 
\end{tabular}
\label{table:nonlin} 
\end{table}

\begin{center}
\textbf{V. Discussion}
\end{center}

Structural distortion is common among ABO$_{3}$ perovskites. The solidity of the cubic structure in this family is often quantified by a tolerance factor:
\begin{equation}
t = (r_{A}+r_{O})/\sqrt{2}(r_{B}+r_{O})
\end{equation}

Here, r$_{A,B,O}$ represent the atomic radii. In order to have a perfect cube, on needs to keep t=1. Now, SrTiO$_{3}$ has a tolerance factor close to unity (t=1.009); This is much lower than in BaTiO$_{3}$(t=1.07) and significantly larger than in CaTiO$_{3}$ (t=0.97)\cite{Zhong}. Unsurprisingly, SrTiO$_{3}$ keeps its cubic structure down to 105 K, contrary to CaTiO$_{3}$ and BaTiO$_{3}$, which are both distorted well above room temperature.

The distortion specific to SrTiO$_{3}$, the clockwise-counterclockwise rotation of adjacent octahedra in a single (001) plane (dubbed antiferrodistortive), is one of the possible twenty-three octahedra tilts in the ABO$_{3}$ system, which were first classified by Glazer\cite{Glazer} and subsequently discussed by other authors\cite{Woodward,Thomas}.

Another common distortion is the loss of inversion symmetry as a consequence of a displacement of atoms A and/or B  from the center of respective polyhedra. This distortion leads to ferroelectricity. It has been commonly believed that these two distortions compete with each other, since there is no octahedral tilt in either of the two celebrated ferroelectrics of the family, BaTiO$_{3}$ and PbTiO$_{3}$. On the other hand, a tilt of octahedra, reminiscent of  the one seen in SrTiO$_{3}$ below 105 K, has been recently found in non-ferroelectric (and magnetic) EuTiO$_{3}$ below a transition temperature of  280~K\cite{Bussmann}.

Our first result is a counterexample to the competitive relation assumed between the two instabilities. We find that in the case of lightly-doped Ca:SrTiO$_{3}$, the increase in Ca content strengthens both instabilities. It amplifies the angle of octahedra rotation and enhances the ferroelectric transition temperature. Let us put this result in the context of theoretical predictions.

Zhong and Vanderbilt theoretically addressed the competition between the two distortions in SrTiO$_{3}$\cite{Zhong}. They computed the fate of the system subject to negative pressure and found that an expansion of the lattice parameter leads to the replacement of the antiferrodistortive transition by a ferroelectric distortion. They found a window in which both stabilities are present, but their pressure dependence is opposite to each other.  More recently, in another theoretical treatment of the problem, Aschauer and Spaldin found that while for small angles of octahedral rotation ferroelectricity is indeed suppressed, larger octahedral tilt angles can stabilize ferroelectricity\cite{Aschauer}.

As a consequence of the smaller ionic radius of calcium compared to strontium, calcium doping can be considered as a case of applying negative chemical pressure. However, infinitesimal calcium substitution generates also an anisotropic strain field\cite{Kleemann1}, which has not been taken into account in these theoretical treatments. Therefore, our case cannot be directly compared to the theoretical predictions for stoichiometric systems. However, the tilt enhancement accompanying the emergence of ferroelectricity indicates a trend, which may be compatible with what is theoretically predicted by Aschauer and Spaldin\cite{Aschauer}.

The increase in the transition temperature induced by calcium doping found here is to be contrasted with the effect of oxygen reduction, documented by studies of ultrasound attenuation\cite{Bauerle} and neutron scattering\cite{Hastings}. Both studies found that oxygen removal  and introducing $10^{20} $cm$^{-3}$ electrons shifts the transition downward by  of 20 K.
\begin{figure*}\centering
\resizebox{!}{0.55\textwidth}
{\includegraphics{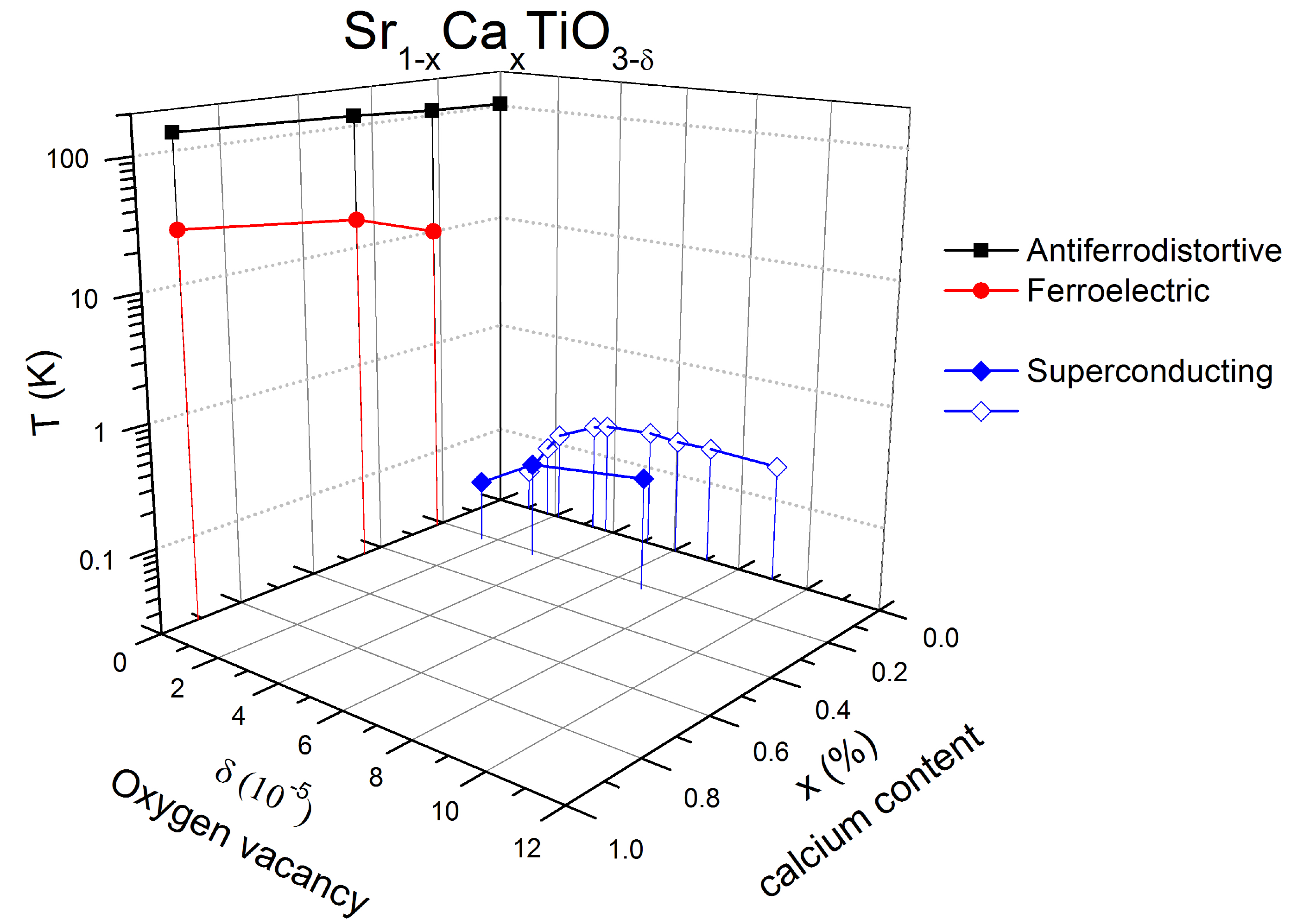}}
\caption{Three-dimensional phase diagram of Sr$_{1-x}$Ca$_{x}$TiO$_{3-\delta}$. The three transition temperatures evolve as a function of calcium substitution (x) and oxygen vacancies ($\delta$). }
\end{figure*}
The second outcome of this study is that one can introduce metallicity and superconductivity by removing a slight fraction of oxygens in Ca:SrTiO$_{3}$. Starting from a ferroelectric, we find excellent metallicity with a carrier concentration in the range of 10$^{17}$cm$^{-3}$. This is to be contrasted with the robustness  of the insulating state in ferroelectric BaTiO$_{3}$\cite{Kolodiazhnyi}. There, one needs to reach a carrier concentration of $1.6\times 10^{20} $cm$^{-3}$ to cross the insulator-to-metal boundary. Moreover, the metal found there is too dirty to present a sharp Fermi surface and, of course, there is no superconducting transition. One important clue is that the zero-temperature dielectric coefficient is much larger in SrTiO$_{3}$ than in BaTiO$_{3}$. Therefore, the Bohr radius is much longer in the former system pulling down the threshold of metal-insulator transition.

The observation of superconductivity in oxygen-deficient in slightly doped Ca:SrTiO$_{3}$ is a first step to establish the precise boundaries of the phase space occupied by this intriguing superconducting state. There are theoretical proposals linking this superconductivity with either of the two instabilities. According to the one proposed by Appel\cite{Appel}, a crucial role in the formation of the Cooper pairs is played by a soft phonon mode associated with octahedra rotation. Rowley and co-workers have recently invoked the possible role of the quantum fluctuations of the ferroelectric order in causing superconductivity\cite{Rowley}. Finally, the weakening of Coulomb repulsion due to strong screening offered by the large dielectric coefficient plays an important role in Takada's proposed route towards superconductivity\cite{Takada}.

A recent study\cite{Lin3} has found that in SrTiO$_{3-\delta}$ bulk superconductivity occurs at a temperature significantly lower than what is detected by resistivity. Therefore, the broad resistive superconducting transitions seen in our sample should be treated with caution. However, it appears clearly that superconductivity is somewhat weakened by calcium substitution. One can invoke two possible explanations for this.

The first possibility would be that this is an issue in material science. If oxygen vacancies happen to cluster around calcium sites, the inhomogeneity in the oxygen-reduced and calcium-doped samples will be larger. In other words, reduced homogeneity may be a result of spatial correlation between theses two types of defects introduced to the lattice. We note that calcium substitution, in addition to decreasing T$_{c}$ and broadening the transition, lowers the carrier mobility in the normal state. Such a reduced mean-free-path is indeed what is expected in the case of enhanced spatial inhomogeneity. Another and more speculative line of  thought would invoke the loss of the inversion symmetry in a calcium-substituted site. This may have microscopic consequences for the formation of Cooper pairs. Non-centrosymmetric superconductors have attracted much attention during the last decade\cite{Bauer}. The partial absence of inversion center in our Ca-doped superconductor is reminiscent of such superconductors.

It is instructive to construct a three-dimensional phase diagram presenting the evolution of the three transition temperatures as a function of $x$ and $\delta$ in  Sr$_{1-x}$Ca$_{x}$TiO$_{3-\delta}$. Assuming that two mobile carriers are introduced by one oxygen vacancy, our data leads to Fig. 9. As seen in the figure, superconducting and ferroelectric ground states are not immediate neighbours, which is not surprising as superconductivity is an instability of metallic state and ferroelectricity an instability in the insulating state. This is to be contrasted with the phase diagram of those superconducting families in which superconducting and magnetic ground states share a common border. The exploration of the region below x=0.0022 would shed more light on the cooperative or competitive relation between superconductivity and the two other instabilities.

\begin{center}
\textbf{V. Conclusion and summary}
\end{center}

Substituting strontium with calcium enhances the transition temperature of the antiferrodistortive transition and induces ferroelectricity in SrTiO$_{3}$. oxygen-deficient-calcium doped strontium titanate is a dilute metal with mobile carriers giving rise to quantum oscillations. This metal undergoes a superconducting transition a slightly lower $T_{c}$ and broader transitions than in SrTiO$_{3-\delta}$. These two findings establish this system as an unusual case of proximity between these three instabilities.

Monitoring the survival of superconductivity in this context is a first step to clarify the possible link between the superconducting order and the two neighboring lattice instabilities. Further studies are needed to map this region of the phase diagram in more detail and to clarify the origin of the broadened superconducting transition.

\begin{center}
\textbf{Acknowledgements}
\end{center}

This work was supported in Brazil by the FAPESP (2009/54001-2 AND 2010/06637-2), CNPq (308162/2013-7), PRP-USP, CAPES and FAPEMIG. In France it was supported by the QUANTHERM and SUPERFIELD projects funded by Agence Nationale de Recherche. In Germany it was supported by DFG research grant HE-3219/2-1 as well as by the Institutional Strategy of the University of Cologne within the German Excellence Initiative. K.B. acknowledges University of Sa\~o Paulo for a visiting professorship during which this work was initiated.

\end{document}